\documentclass[multphys,vecphys]{svmult}

\usepackage{makeidx}         % allows index generation
\usepackage{graphicx}        % standard LaTeX graphics tool
\usepackage{multicol}        % used for the two-column index
\usepackage[bottom]{footmisc}% places footnotes at page bottom
\makeindex             % used for the subject index
\begin{document}

\title*{2dF--SDSS LRG and QSO (2SLAQ) Survey: Evolution of the Most Massive Galaxies}
\titlerunning{2SLAQ: Evolution of the Most Massive Galaxies}
\author{Robert C, Nichol,\inst{1} Russell Cannon,\inst{2} Isaac Roseboom,\inst{3} David Wake\inst{4}\, for the 2SLAQ Collaboration}

\institute{ICG, University of Portsmouth, Portsmouth, PO1 2EG, UK
\and Anglo-Australian Observatory, PO Box 296, Epping, NSW 1710, Australia
\and Dept. of Physics, University of Queensland, QLD 4072, Australia
\and Dept. of Physics, Durham University, South Road, Durham DH1 3LE, UK}
\maketitle

{\bf Abstract:} The 2dF--SDSS LRG and QSO (2SLAQ) survey is a new survey of distant Luminous Red Galaxies (LRGs) and faint quasars selected from the Sloan Digital Sky Survey (SDSS) multi--color photometric data and spectroscopically observed using the 2dF instrument on the Anglo-Australian Telescope (AAT). 
In total, the 2SLAQ survey has measured over 11000 LRG redshifts, covering $180 {\rm deg^2}$ of SDSS imaging data, from 87 allocated nights of AAT time.  Over 90\% of these galaxies are within
the range $0.45<z<0.7$ and have luminosities consistent with $\ge 3L^{\star}$. When combined with the lower redshift SDSS LRGs, the evolution in the luminosity function of these LRGs is fully consistent with that expected from a simple passive (luminosity) evolution model. This observation suggests that at least half of the LRGs seen at $z\simeq0.2$ must already have more than half their stellar mass in place by $z\simeq0.6$, i.e., our observations are inconsistent with a majority of LRGs experiencing a major merger in the last 6 Gyrs. However, some ``frosting" (i.e., minor mergers) has taken place with $\sim5\%$ of LRGs showing some evidence of recent and/or on--going star--formation, but it only contributes $\sim1\%$ of their stellar mass.

\section{2SLAQ Survey}
\label{2slaqsurvey}

\begin{figure}[t]
\centering
\includegraphics[height=4in,angle=270]{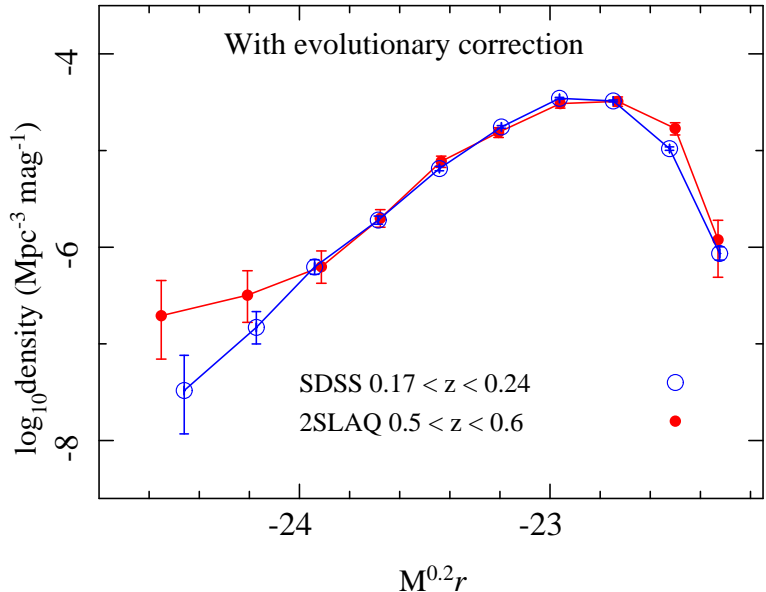}
\caption{\label{lumfn} The M$_{^{0.2}r}$ (absolute magnitude at $z=0.2$) luminosity function after passive evolution corrections for both the SDSS (open data points) and 2SLAQ (solid data points) LRG samples have been made. The points are plotted with their one sigma error bars as described in Wake et al. (2006)}
\end{figure}

\begin{figure}[t]
\centering
\includegraphics[height=3.5in]{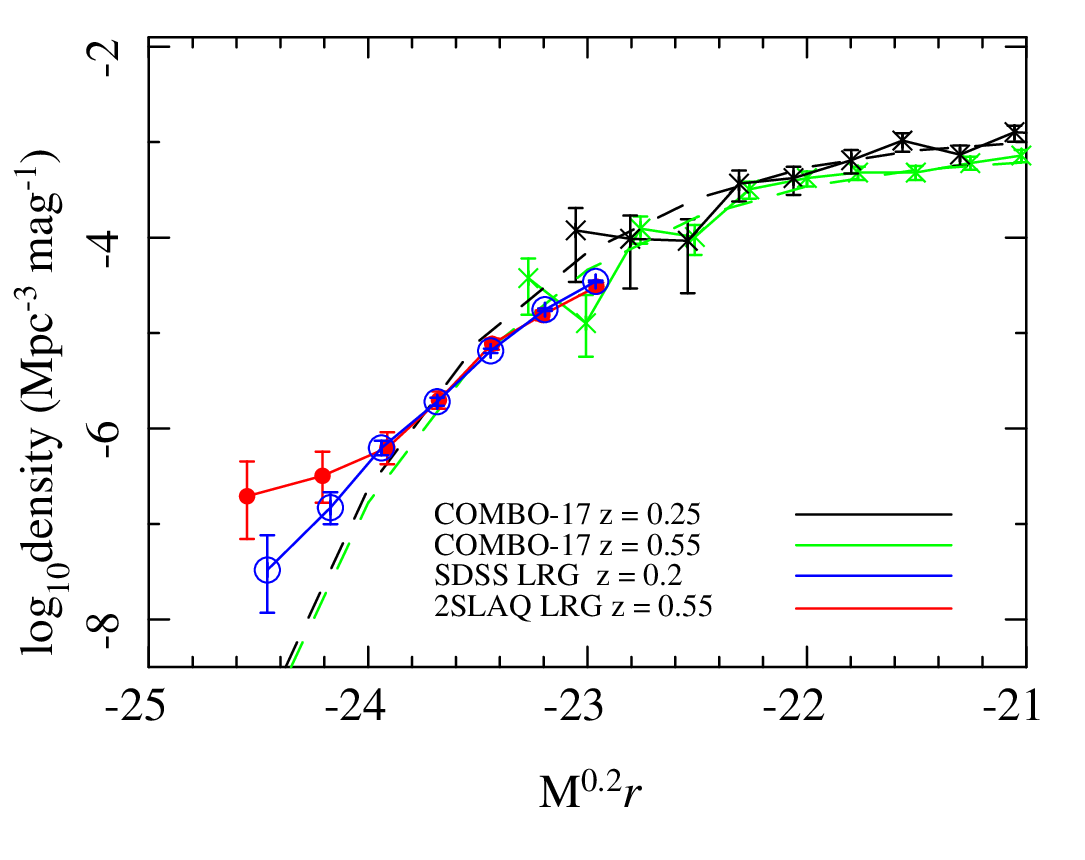}
\caption{\label{lumfncombo} The M$_{^{0.2}r}$ luminosity function with passive evolution corrections for the SDSS (blue open data points), 2SLAQ (red solid data points) LRG samples, and the COMBO-17 red galaxies at z = 0.25 (black open stars) and z = 0.55 (green solid stars). The dashed lines show the Schechter function fit to the COMBO-17 points. The points are plotted with their one sigma errors.}
\end{figure}

\begin{figure}[t]
\centering
\includegraphics[scale=0.5,angle=270]{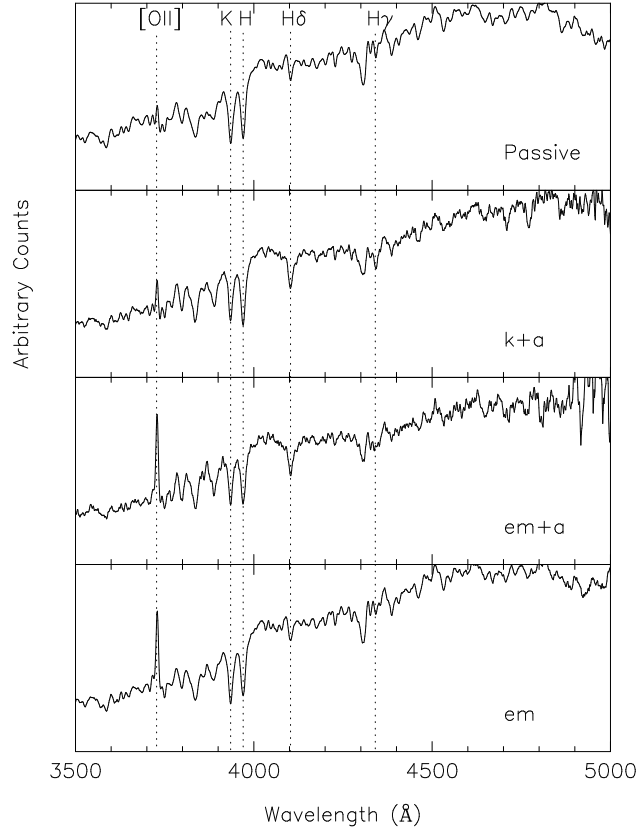}
%
% If not, use
%\picplace{5cm}{2cm} % Give the correct figure height and width in cm
%
\caption{The combined spectrum for the four spectral classifications comsidered in Roseboom et al. (2006) i.e. passive (top), ``k+a", emission ([OII]) plus absorption (H$\delta$), and then just emission ([OII]) on the bottom. Note the increase in the absorption strength of the higher order Balmer lines (H$\gamma$,H$\epsilon$, etc) in the middle two combined spectra.}
\label{combspec}
\end{figure}

The 2dF--SDSS LRG and QSO (2SLAQ) survey is a collaboration of over 70 UK, US and Australian astronomers focused on studying the evolution with redshift of Luminous Red Galaxies (LRGs) and faint quasars. The details of the 2SLAQ survey can be found in a series of recent papers, including Cannon et al. (2006) and Richards et al. (2005). In this paper, we focus on two recent works on the evolution of LRGs with redshift (Wake et al. 2006; Roseboom et al. 2006) from the joint SDSS and 2SLAQ surveys.

\section{Evolution of the Luminosity Function}

As seen in Figure \ref{lumfn}, the SDSS and 2SLAQ luminosity functions (LFs) brighter
than M$_{^{0.2}r}$ = -22.6 are in excellent agreement when
the passive evolution corrections are included. 
The agreement of these luminosity functions is further confirmed by
calculating the integrated number and luminosity density of LRGs brighter than M$_{^{0.2}r}$ = -22.5, which  to
better than 10\% out to $z=0.6$. Throughout the analysis, 
the same simple passive evolution model was used for predicting and correcting
the colours and luminosities of LRGs as a function of redshift, and
this agreement demonstrates the lack of any extra evolution, beyond
the passive fading of old stars, out to $z \simeq 0.6$. 

It may appear that our lack of extra evolution beyond passive (out to
$z\sim0.6$) is in conflict with recent results from the COMBO-17 (C17) and DEEP2 (Bell et al. 2004; faber et al. 2005). These smaller--area, but deeper (in magnitude limit and redshift),
surveys find evidence for a change in the density of red galaxies out
to $z\sim1$ beyond that expected from passive fading of the stellar
populations. For example, Faber et al. (2005) report a quadrupling of
$\phi^*$ for red galaxies since $z=1$, although this result is
strongest in their highest redshift bin, where they admit their data
are weakest. A direct comparison with these deeper surveys is
difficult because of the differences in colour selections used for the
surveys, as well as the relative luminosity ranges probed by the
different surveys, i.e., the 2SLAQ survey is designed to probe
galaxies brighter than a few $L^*$, while the DEEP2 and C17
surveys effectively probe galaxies below $L^*$ at $z\sim0.6$ (due to
their smaller areal coverage and fainter magnitude limits).

However, to facilitate such a comparison, we show in Figure
\ref{lumfncombo}, the LFs from Figure
\ref{lumfn}, and the C17 red galaxy LFs for the same redshift range and K+e corrected
to M$_{^{0.2}r}$. We only plot our LFs to M$_{^{0.2}r} <$ -22.9 as we do 
not include all the red galaxies fainter than this due to the SDSS LRG 
selection criteria.
 Figure \ref{lumfncombo} demonstrates that when 
one restricts the
data to the same redshift range, there is excellent qualitative
agreement between the 2SLAQ and C17 luminosity functions. We are unable to 
make a quantitative comparison due to the difficulty in exactly matching the 
selection criteria of the two surveys.
 Taken
together, the surveys shown in Figure \ref{lumfncombo} extend the 
evidence for no evolution in the LF of
LRGs to M$_{^{0.2}r} < -21$, which is close to $L^{\star}$ in the LF.
Figure \ref{lumfncombo} also demonstrates that these two surveys 
are probing
different luminosity regimes at $z<0.6$ as there is at most only 0.5
magnitudes of overlap in their LFs in which the C17 survey is becoming seriously affected 
by small number statistics due to its smaller areal coverage.

The luminosity functions given in Figures \ref{lumfn} and \ref{lumfncombo} place tight
constraints on models of massive galaxy formation and evolution. Our
results appear to favour little, or no, density evolution, i.e., there are already enough LRGs
per unit volume at $z\simeq0.6$ to account for the density of LRGs
measured at $z\simeq 0.2$. Using a simple model for "dry (major) mergers", we find that the
2SLAQ and SDSS LFs are
consistent with each other without any need for merging. At the
3$\sigma$ level, we can exclude merger rates of $>50\%$, i.e., more
than half the LRGs at $z=0.2$ are already well-assembled, with more
than half their stellar mass in place, by $z\simeq0.6$. 
Our limit is barely consistent with the predictions in Figure 5 of
De Lucia et al. (2006), where they show that $\sim 50\%$ of $z=0$
massive ellipticals have accreted 50\% of their stellar mass since
$z\simeq 0.8$. 

\section{Minor Mergers}

Our simple model for ``dry (major) mergers" above does not constrain the
rate of minor mergers involving LRGs. For example, Roseboom et al. (2006) 
determined the recent star formation histories of the SDSS and 2SLAQ LRGs based on the H$\delta$ and [OII] lines (see Figure \ref{combspec} for examples). While the majority ($>$80 per cent) of LRGs show the spectral properties of an old, passively evolving, stellar population, a significant number of LRGs show evidence for recent and/or ongoing star formation in the form of ``k+a'' (2.7 per cent), emission+absorption (1.2 per cent) or just [OII] emission LRGs (8.6 per cent). Therefore, $\sim5\%$ of LRGs have evidence for recent star--formation (k+a's and the emission+absorption), while it is unclear what percentage of the [OII] emitters for due to on--going star--formation or have an AGN. Furthermore, the [OII] emitters could be interlopers in the sample, scattered over the color boundaries. 

By dividing the sample into 2 redshift subsamples from $0.45<z<0.55$ and $0.55<z<0.65$, and comparing to a $z\sim0.15$ sample selected from SDSS, it is observed that the fraction of ``k+a" LRGs increases with redshift as $(1+z)^{2.8\pm0.7}$. Spectral synthesis models suggest that these LRGs could originate from passive LRGs which have undergone 
a starburst involving only $\sim 1\%$ of their total stellar mass. Therefore, the $\simeq5\%$ of LRGs that show evidence of recent and/or on--going star--formation (k+a and emission+absorption galaxies) are probably produced by minor mergers with gas--rich dwarf galaxies and represent a "frosting" of new stars on top of the majority older stellar population.

RCN thanks Hans Boehringer for still including this paper in the proceedings even though he missed the conference due to the August $10^{th}$ 2006 terrorist alert at Heathrow Airport. We all thank the 2SLAQ team for their hard work and allowing us to present their work here. We again thank the AAO staff for all their assistance during the collection of the 2SLAQ data. 

%
% BibTeX users please use
% \bibliographystyle{}
% \bibliography{}

\begin{thebibliography}{99.}
\bibitem{2004ApJ...608..752B} Bell, E.~F., et al.\ 2004, ApJ, \textbf{608}, 752 
\bibitem{2006MNRAS.372..425C} Cannon, R., et al.\ 2006, MNRAS, \textbf{372}, 425 
\bibitem{2006MNRAS.366..499D} De Lucia, L., et al.\ 2006, MNRAS, \textbf{366}, 499
\bibitem{Faber} Faber, S., et al., 2005, astro-ph/0506044
\bibitem{2005MNRAS.360..839R} Richards, G.~T., et al.\ 2005, MNRAS, \textbf{360}, 839 
\bibitem{2006MNRAS.373..349R} Roseboom, I.~G., et  al.\ 2006, MNRAS, \textbf{373}, 349 
\bibitem{2006MNRAS.372..537W} Wake, D.~A., et al.\ 2006, MNRAS, \textbf{372}, 537
\end{thebibliography}
%
% Non-BibTeX users please use

%%%%%%%%%%%%%%%%%%%%%%%%%%%%%%%%%%%%%%%%%%%%%%%%%%%%%%%%%%%%%%%%%%%%%%  }

%%%%%%%%%%%%%%%%%%%%%%%%%%%%%%%%%%%%%%%%%%%%%%%%%%%%%%%%%%%%%%%%%%%%%%

\printindex
\end{document}